\documentclass[preprint2]{aastex}

\usepackage{graphicx}
\usepackage{rotating}
\usepackage{natbib}
\usepackage{paralist}
\usepackage{psfrag,color}
\usepackage[latin1]{inputenc}

\newcommand{\lam}{$\lambda$}
\newcommand{\hei}{He\,{\sc i}}
\newcommand{\heii}{He\,{\sc ii}}
\newcommand{\cy}{Cyg\,OB2\,\#9}

\newcommand{\kms}{km\,s$^{-1}$}
\newcommand{\xmm}{{\sc XMM}\emph{-Newton}}

\shorttitle{First orbital solution for \cy}
\shortauthors{Naz\'e et al.}

\begin{document}

\title{First orbital solution for the non-thermal emitter \cy\altaffilmark{1}}

\author{Y. Naz\'e,\altaffilmark{2,3} 
  Y. Damerdji,\altaffilmark{2}
  G. Rauw,\altaffilmark{2,3}
  D.C. Kiminki,\altaffilmark{4}
  L. Mahy,\altaffilmark{2}
  H.A. Kobulnicky,\altaffilmark{4}
  T. Morel,\altaffilmark{2}
  M. De Becker,\altaffilmark{2,5}
  P. Eenens,\altaffilmark{6}
  and C. Barbieri\altaffilmark{7}}
\email{naze@astro.ulg.ac.be}


\altaffiltext{1}{Based on observations collected at the Haute-Provence Observatory and with XMM-Newton, an ESA Science Mission with instruments and contributions directly funded by ESA Member States and the USA (NASA).}
\altaffiltext{2}{GAPHE, D\'epartement AGO, Universit\'e de Li\`ege, All\'ee du 6 Ao\^ut 17, Bat. B5C, B4000-Li\`ege, Belgium ; email: naze@astro.ulg.ac.be}
\altaffiltext{3}{Research Associate FRS-FNRS}
\altaffiltext{4}{Department of Physics \& Astronomy, University of Wyoming, Laramie, WY 82070, USA}
\altaffiltext{5}{Observatoire de Haute-Provence, F-04870 St.Michel l'Observatoire, France}
\altaffiltext{6}{Departamento de Astronomia, Universidad de Guanajuato, Apartado 144, 36000 Guanajuato, GTO, Mexico}
\altaffiltext{7}{Dipartimento di Astronomia, Universit\'a degli studi di Padova, vicolo Osservatorio 2, 35122 Padova, Italy}

\begin{abstract}
After the first detection of its binary nature, the spectroscopic monitoring of the non-thermal radio emitter \cy\ (P=2.4\,yrs) has continued, doubling the number of available spectra of the star. Since the discovery paper of 2008, a second periastron passage has occurred in February 2009. Using a variety of techniques, the radial velocities could be estimated and a first, preliminary orbital solution was derived from the \hei\,\lam\,5876\AA\ line. The mass ratio appears close to unity and the eccentricity is large, 0.7--0.75. X-ray data from 2004 and 2007 are also analyzed in quest of peculiarities linked to binarity. The observations reveal no large overluminosity nor strong hardness, but it must be noted that the high-energy data were taken after the periastron passage, at a time where colliding wind emission may be low. Some unusual X-ray variability is however detected, with a 10\%\ flux decrease between 2004 and 2007. To clarify their origin and find a more obvious signature of the wind-wind collision, additional data, taken at periastron and close to it, are needed.
\end{abstract}

\keywords{binaries: spectroscopic -- stars: early-type -- stars: emission-line -- stars: individual (\cy) -- X-rays: stars}

\section{Introduction}
As the main responsibles for mechanical input, chemical enrichment and ionizing radiation in galaxies, massive stars (i.e., with masses $>$10\,M$_{\odot}$) are important objects of the stellar population. However, these objects are rare: in the Galaxy, about 300 Wolf-Rayet stars (i.e., fewer than exoplanets) are known \citep{van06}, as well as about 400 O-type stars (plus 700 candidates that have sometimes be classified as O, see \citealt{sot08}). A natural consequence is that many aspects of these stars remain poorly known, even in the case of their major property, the stellar wind, for which ``basic'' quantities such as the mass-loss rates are still heavily debated at the present time \citep[e.g.][]{sun10}. 

In this context, non-thermal radio emitters associated with massive stars form an even more limited group, with fewer than 40 cases known in our Galaxy \citep{deb07,ben09}, but these objects can provide unique insights into the physics of stellar winds since both phenomena are intimately linked. 

Observing such non-thermal radio emission implies two pre-requisites: the presence of both a magnetic field and a population of relativistic electrons. Direct detection of the former is notoriously difficult in massive stars, which display only few broad lines and therefore have a weak Zeeman signature. However, as some of the descendants of massive stars (i.e., neutron stars) are clearly magnetic, there was little doubt that magnetic fields are not totally absent in these objects. Indeed, the last decade saw the collection of the first evidences of magnetism in massive stars thanks to spectropolarimetric monitorings \citep[e.g., ][ with even a tentative detection in a non-thermal emitter in \citealt{hub08}]{don02}. 

The second requirement, the population of relativistic electrons, is ultimately linked to the presence of shocks in the ionized winds \citep[for a review see][]{deb07}. Acceleration then takes place through the first-order Fermi mechanism, also called ``diffusive shock acceleration'', where electrons iteratively gain energy by crossing several times the shock zone. Shocks are not ubiquitous in stellar winds: they can either be intrinsic to the winds themselves (as the line-driven mechanism producing the wind is intrinsically unstable) or arise in massive binaries from the collision of the two stellar winds. Recent theoretical modelling showed that only the latter can account for the observed non-thermal radio emission \citep{van05}. Indeed, the large majority of the known cases of non-thermal emitters are confirmed or suspected binaries - only 3 of the 17 WRs and 3 of the 16 O stars listed by \citet{deb07} totally lack evidence for binarity. However, this can often be explained by an inadequate monitoring, or sometimes simply the absence of any monitoring. Once an adequate observing campaign is organized, proofs of binarity are often found, as exemplified by our recent successes in this domain (e.g., Cyg\,OB2\,\#8A, \citealt{deb04}, and 9\,Sgr \citealt{rau05b}, 2010, in prep) which greatly improved the multiplicity census of non-thermal radio emitters associated with O-type stars. Among the remaining objects without evidence of multiplicity, the early-type star \cy\ (O5I) is clearly a target of choice.

Located in the rich association Cyg OB2 \citep{kno00}, \cy\ was one of the first O-stars shown to be a non-thermal radio-emitter \citep{abb84}. The presence of this emission remained problematic for years. The first direct evidence of the binary nature of \cy\ was only reported in 2008 thanks to a dedicated, long-term spectroscopic monitoring \citep{naz08}. In the same year, \citet{van08} revealed a long-term modulation of the radio emission and derived for it a period of 2.355\,yrs, interpreted as being associated with the binary orbit. This radio flux appeared minimum when the stellar lines were unblended in the optical spectrum. Up to now, however, a full orbital solution and a detailed modelling of the system are still missing. This paper aims at beginning to fill this gap, by tackling the first problem: the derivation of the orbital parameters.

It must be recalled that \cy\ is not an easy star to analyze. On the one hand, it is strongly extinguished ($E[B-V]=2.11$, \citealt{mas91}) with two consequences. First, the spectrum of \cy\ displays very strong interstellar lines, often affecting the stellar lines, even those usually quite uncontaminated (e.g., C\,{\sc iv}\,\lam\,5812). Second, the star appears faint, especially at blue wavelengths where most lines used for spectral classification are found. On the other hand, the stellar lines remain totally blended for about 80\% of the orbit ; a clear doubling of the lines can only be seen during a few months of the 2.355\,yrs period. This is due to the large eccentricity of the system but also to the width of the lines ($FWHM_{\rm HeI\lambda5876}\sim3$\,\AA). 

Despite these difficulties, we have continued our monitoring of \cy, with the hope of improving our knowledge of this system. This paper reports on the new data collected since January 2008, including at orbital phases close to the periastron passage of 2009 (the periastron itself was unobservable as \cy\ was in conjunction with the Sun at that time). A preliminary orbital solution is presented here for the first time. This is a necessary and crucial step towards the full modelling of the system and the derivation of the winds' parameters.

Complementary high-energy data are also presented here, as they directly relate to the question of colliding winds. At first, one could expect for these peculiar objects a non-thermal X-ray emission, a direct companion to that observed in the radio range. However, direct and undisputed evidence of such emission is still lacking: its detection awaits the advent of sensitive observatories in the $>$10\,keV range \citep{deb09}. In the meantime, one could however try to find evidence for wind-wind collision in the {\it thermal} X-ray emission which dominates the 0.3--10.\,keV range, the preferred bandwidth of the current sensitive facilities (\xmm, {\it Chandra}). Indeed, some binaries display strong wind-wind shocks which are able to produce very hot plasma, hence hard and bright X-rays. Moreover, in these cases, the X-ray emission is modulated as the stars orbit each other \citep[for a review see][]{gud09}. It is therefore important to check whether the X-ray emission of \cy\ bears the signature of wind-wind collision. This would yield a further proof of the link between binarity and non-thermal radio emission, as well as additional constraints on the winds' parameters.

The paper is organized as follows: Sections 2 and 3 present the optical dataset and its analysis, Section 4 details the X-ray properties of \cy, and Section 5 summarizes our results.

\section{The optical dataset}
The observations obtained until early January 2008 have been described in length in \citet{naz08}, and only the new data will be described here. These new observations were obtained at the Haute-Provence observatory (OHP, France) and the Wyoming Infrared Observatory (WIRO, USA); an archival dataset taken at San Pedro M\'artir (SPM, Mexico) in 2004 has been added to the new analysis. A journal of these observations is provided in Table \ref{journ}. 

At the OHP, six additional spectra in the yellow range and one in the blue range were obtained in 2008 using the 1.52m telescope equipped with the Aur\'elie spectrograph (grating \#3, $R\sim9000$). In 2009, the Sophie \'echelle instrument ($R=35000$, 39 orders over the domain $3900-6900$\AA) installed on the 1.93m telescope observed the system 9 times in the high-efficiency mode. For each dataset, the typical exposure time was 1800--7200s ; the observations were sometimes split into several individual exposures which were finally combined if taken within 1--15\,days. The data were first reduced in a standard way, smoothed by a moving box average and finally normalized. Note that data from August 10 and 13 2009 were unusable due to a technical problem.

In Wyoming, the observations were obtained with the 2.3m telescope equipped with the WIRO-Longslit spectrograph (1800~l~mm$^{-1}$ grating in first order, $R\sim4000$). Exposure times varied from 720 to 5400~s (generally in multiples of 600--900~s) depending on weather conditions. All datasets were reduced using standard IRAF reduction routines as outlined in \citet{kim07}.

One archival \'echelle spectrum, taken with the 2.1m telescope of SPM equipped with the Espresso spectrograph ($R=18000$, 27 orders in over the range $3780-6950$\AA), was also made available and added to our dataset. These data were reduced in a standard way using MIDAS.

To improve the wavelength calibration, we took advantage of the high reddening and used several narrow, well-marked Diffuse Interstellar Bands (DIBs) close to major spectral lines. Their velocity shifts relative to a chosen reference dataset (the \'echelle spectra taken by Sophie in October 2007) were measured using a cross-correlation method. The measured radial velocities (RVs) of the stellar lines (see below) were then corrected by the shifts derived from a close DIB (e.g., DIB\,\lam\,5780\AA\ for \hei\,\lam\,5876\AA). This ensures that the wavelength calibration is correct to within 5--10\,\kms\ in the worst case.

In total, our dataset now comprises 20 \'echelle spectra (Sophie at OHP, AFOSC at Asiago, Espresso at SPM), 8 long-slit spectra at red wavelengths (Aur\'elie at OHP, Asiago), 10 yellow spectra (Aur\'elie at OHP), 1 blue spectrum (Aur\'elie at OHP), and 4 yellow-to-red, low-resolution spectra (WIRO). These data were taken between 2003 and 2009, with a more intense monitoring since 2006.

\section{Towards a first orbital solution}

\subsection{Radial velocities}

To derive an orbital solution, it is necessary to (1) choose adequate stellar lines and (2) secure good estimates of the radial velocities (RVs). 

Choosing reliable stellar lines is no easy task for \cy. With strong interstellar lines and low signal-to-noise ratios, many lines have to be discarded. While the splitting is seen near periastron for \hei\,\lam\,4471 and \heii\,\lam\,4542, the strong noise in the blue domain prevents us from deriving good RV estimates from these lines throughout the whole orbit. Metal lines, such as C\,{\sc iv}\,\lam\lam\,5801,5812 or O\,{\sc iii}\,\lam\,5592, are strongly contaminated by interstellar lines, rendering their measurement difficult. In addition, the two stars of the system appear so similar in spectral type that there is no obvious, sufficiently strong stellar line belonging to only one component of the pair. Also, the data do not always cover the same wavelength range and the chosen line should indeed belong to the most frequently recorded domain. Our best choice for RV determination is thus \hei\,\lam\,5876, which is simultaneously strong and free from interstellar contamination. However, there is a caveat to this choice: this line is sometimes polluted by emission in some extreme O-type stars. In our dataset, there is no emission above the continuum level, but it is difficult to exclude with 100\% confidence the presence of faint emission which would slightly ``fill'' the photospheric absorptions. Our results are thus clearly preliminary.

\begin{figure}
  \includegraphics[width=8cm]{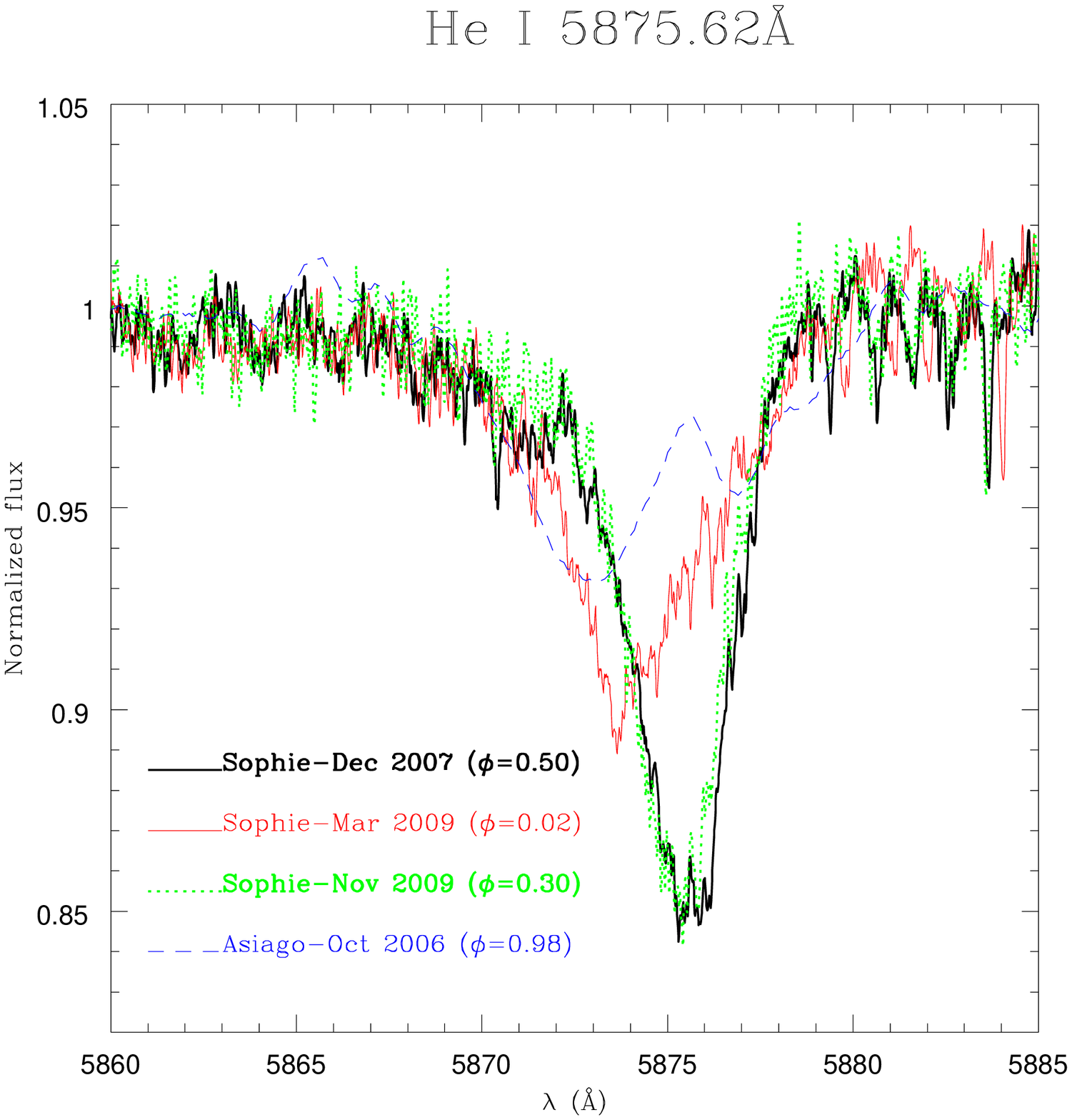}
  \includegraphics[width=10cm, bb= 50 220 540 620, clip]{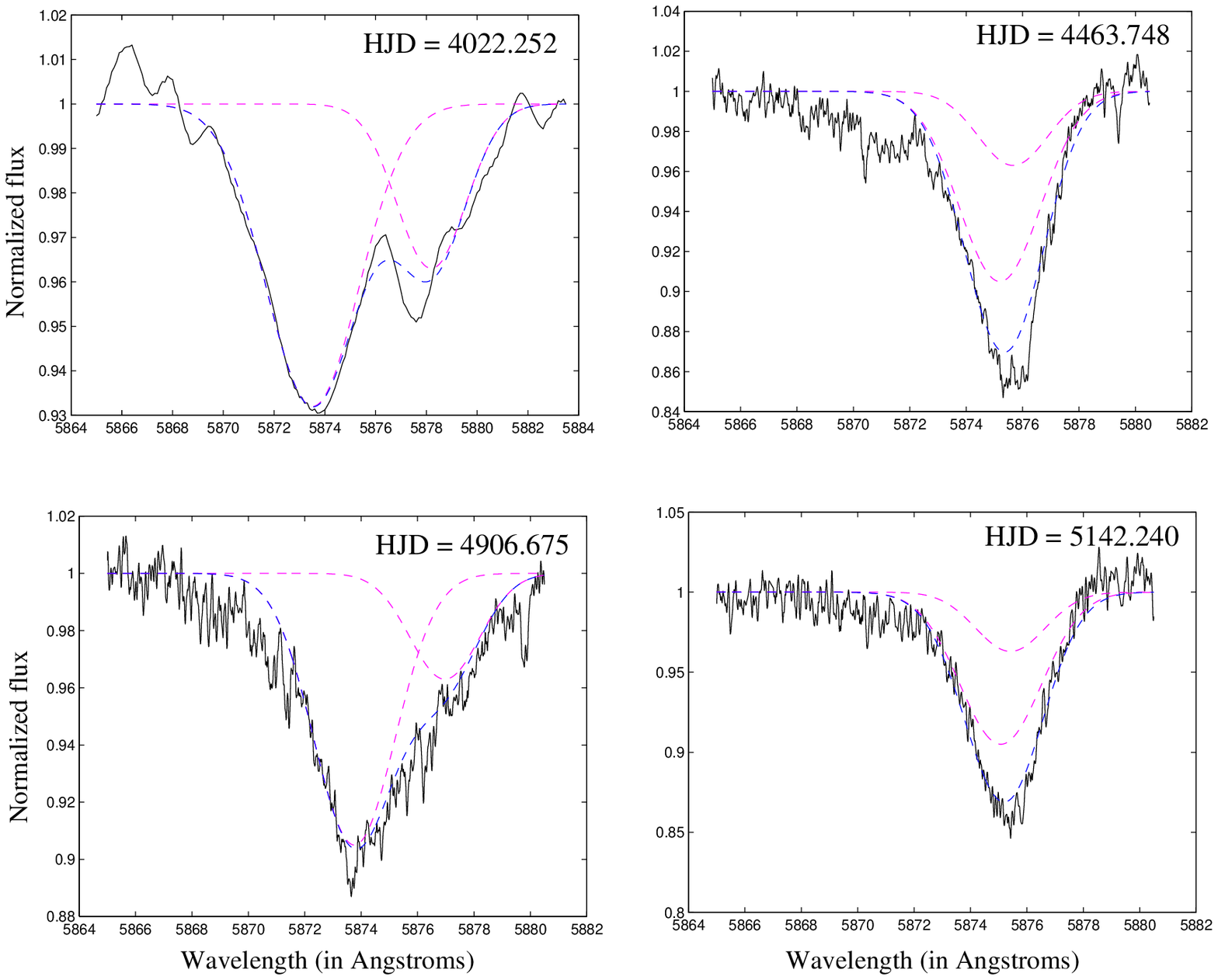}
  \hfill
  \caption{Left: The \hei\,\lam\,5876 line in the Sophie data of $HJD$ 4463.748 (thick solid black line), 4906.675 (thin solid red line, data taken half an orbit later than the previous ones), and 5142.240 (dotted thick green line, data taken a quarter of an orbit later than previous ones), as well as the Asiago data of the 2006 periastron passage (4022.252, dashed blue line). Quoted phases are from the SB2 solution (2nd column of Table \ref{SOLORB}). Right: Deblending using the $\chi^2$ method for these spectra.}
  \label{fig:line}
\end{figure}

\begin{figure}
\centering
  \includegraphics[width=12cm]{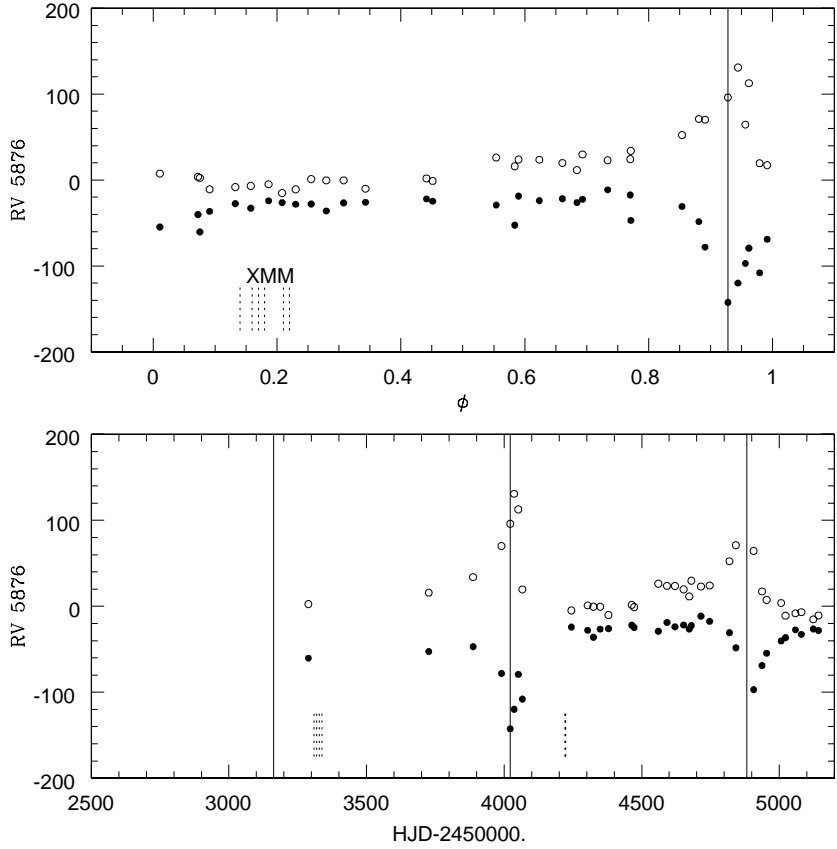}
  \hfill
  \caption{Evolution of the Radial Velocities of \hei\,\lam\,5876 measured by the $\chi^2$ method with time (bottom panel) and phase (top panel, using the ephemeris from \citealt{van08}). The primary (resp. second.) RVs are shown in filled (resp. open) circles\ ; vertical lines indicate 2006 October 13 ($HJD$ 4022.252), the approximate date of the periastron passage, and the dates 2.355yrs before/after. Dotted lines indicate the phases or dates at which the \xmm\ data were taken.}
  \label{fig:RV}
\end{figure}

The stellar lines have first been fitted by Gaussians (usually one, with two only for the 2-3 clearly unblended spectra). The last column of Table \ref{RV} gives the result of a single Gaussian fit to the blended data (i.e., $0.1<\phi<0.9$). However, as the line splitting has clearly been detected \citep{naz08}, at least during some part of the orbit, we tried to improve this RV determination by using the correlation method \citep[TODCOR, see ][]{zuc94}, the disentangling method \citep{gon06,mah10}, and the $\chi^2$-mapping method. Used on a sole line (\hei\,\lam\,5876), the first two methods give poor results. As TODCOR involves the convolution between a typical line shape (i.e., a Gaussian in our case) and the spectrum (reduced here to one single line), the peak of the cross-correlation function is broad, increasing the error bars on the RVs to unacceptable levels. Even at maximum separation, TODCOR does not always yield results within 10\,\kms\ of the simple two-Gaussian fitting, and it was thus discarded. Disentangling the \hei\ line also proved unreliable, the main reason being that due to the orbital configuration, the lines are only partially separated when the primary is blueshifted and the secondary redshifted. The reverse situation, redshifted primary and blueshifted secondary, is never observed (Fig. \ref{fig:line}, left). Therefore, the only solution for fitting two components in \hei\,\lam\,5876 was first to fit each component at maximum separations using Gaussians and then to shift these two Gaussians, keeping their shape (width and depth) constant, to find the minimum $\chi^2$ for each spectrum (Fig. \ref{fig:line}, right). The RVs determined with this method are listed in the third and fourth columns of Table \ref{RV}. When available and close to maximum separation, results from this method are similar for \heii\,\lam\,6683. Fig. \ref{fig:RV} shows the evolution of the RVs, the two periastron passages can be clearly seen, as well as the relative constancy of the RVs during most of the period. When the two components are blended, the RVs are still rather uncertain, as the separation is then much smaller than the line width.

We have detailed above the numerous caveats concerning our measurements and our data (blending during most of the orbit, possible slight contamination by emission, $\chi^2$-mapping vs single Gaussian fit). Caution should thus apply, but it should also be stressed that these observations and measurements represent the best dataset available for \cy. Deriving an orbital solution using these RVs was thus attempted. 

\subsection{Orbital solution}

SB1 solutions were calculated in several steps. First, the best-fit period was selected using the Fourier method of \citet[see also remarks in \citealt{gos01}]{hmm}. A polynomial fit of the folded RV curve was then calculated and an approached orbital solution was derived from it using the best-fit result amongst solutions calculated using methods by \citet{wol67} and \citet{leh}. The former method relies on the derivation of the Fourier expansion (limited to 2 terms) of the observed RV curve followed by an identification of the coefficients with those issued from a series expansion of the theoretical RV curve corresponding to a Kepler orbit ; the latter method uses the amplitude and various integrations under well defined parts of the RV curve to get a first estimate of the orbital parameters. From that starting point, a refined, final solution was found using a Levenberg-Marquardt minimization on a method adapted to eccentric binaries ($e>0.03$) by \citet{sch10}. A check of the correctness of the orbital solution was made based on the recent algorithm of \citet{zec09} which, by performing the fit in terms of the true anomaly, reduces the non-linearity of the RV curve to 3 free parameters instead of 6 (the other 3 free parameters then being readily derived from the periodogram coefficients) ; only few iterations of the Levenberg-Marquardt method are then needed to reach the final solution, which, in our datasets, always agrees with the result of the first derivation. Errors were estimated using the diagonal of the variance-covariance matrix. The SB1 fitting was performed on the RVs of the primary, on the RVs of the secondary, and on the velocity difference between primary and secondary.

SB2 solutions relied on the main idea of the Li\`ege Orbital Solution Package (LOSP)\footnote{The LOSP package and a preprint describing it (Sana \& Gosset, A\&A, submitted) can be downloaded from http://staff.science.uva.nl/$\sim$hsana/losp.html.}. In this method, the secondary and primary velocities are converted into an equivalent SB1 dataset, using a linear orthogonal regression fit between the velocities of the two components. After this transformation, the equivalent SB1 dataset was fitted as described above. 

In both cases, the period was allowed to vary slightly, since \citet{van08} had a relatively large uncertainty ($P=2.355\pm0.015$\,yrs) and our observations were taken 10 to 13 cycles after the radio data. Indeed, the phase of the Oct. 2006 and Feb. 2009 periastron passages seem to occur at $\phi=0.95$ rather than at 0.0 (Fig. \ref{fig:RV}). However, since we cover only two events of maximum separations, the formal improvement on the period error from our sole dataset is not very large.

\begin{figure}
  \includegraphics[width=9cm, bb=20 180 570 675, clip]{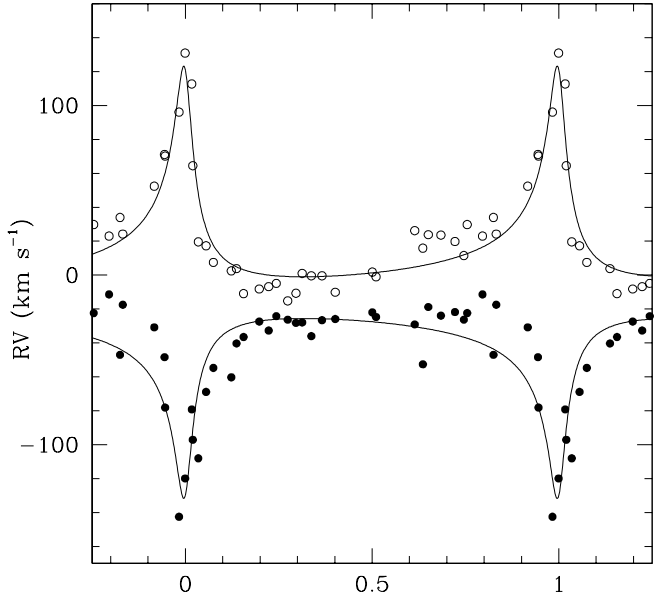}
  \includegraphics[width=8.8cm]{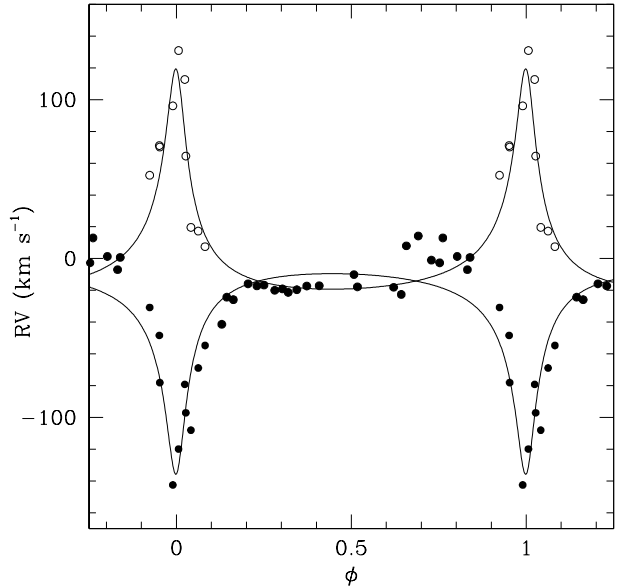}
  \hfill
  \caption{RV curves superimposed on the best SB2 solutions (SB2 shown on the left panel, SB2' on the right panel). The symbols are as before and the errors on the velocities are of the order of 10--20\,\kms.}
  \label{fig:solorb}
\end{figure}

Table \ref{SOLORB} gives the derived orbital parameters and their associated error bars while Fig. \ref{fig:solorb} graphically shows the results of the best SB2 solutions. Note that smaller weights were given to the lower-quality data (0.3 for WIRO and 0.6 for Asiago and SPM, 1 otherwise).

The best orbital solutions rely on our best estimates of the RVs, i.e., those derived from the $\chi^2$-mapping. In each case, the best-fit period is slightly revised downwards but agrees well, within the errors, with the radio determination. As it takes into account both components, the SB1 solution calculated on the RV difference agrees best with the results of the SB2 solution, though the parameters derived for individual SB1 solutions are never at 3-$\sigma$ from the results of the SB2 solution. The mass ratio is close to unity, confirming the similarity of the two stars of \cy\ found by \citet{naz08} who proposed spectral types of O5+O6--7. Using the typical masses of such stars as quoted in \citet{mar05}, the masses derived from the $\chi^2$-mapping imply an inclination of about 45--50$^{\circ}$ if both components are supergiants, or 55-60$^{\circ}$ if both components still belong to the main sequence (which is unlikely in view of the combined O5I spectral type). The eccentricity is large, as previously suspected, with a value of about 0.7--0.75. The velocities of the center-of-mass are quite different for both stars, as could be expected from the fact that the secondary lines are never seen on the blue side. Indeed, our RV curves do not cross (see Fig. \ref{fig:solorb}, left) but, as already mentioned, the RV determination when lines are totally blended is difficult and a slight crossing (by e.g., 10\,\kms) of the true RV curves can therefore not be totally excluded. In any case, it would not change the fact that the center-of-mass velocity of the primary is blueshifted compared to that of the secondary. This is most probably related to the fact that the wind of the primary is stronger than that of the secondary, leading to the formation of the stellar lines of the most extreme star (the primary) not at the photosphere but farther, in the wind itself.

As the RVs are uncertain when the lines are blended, i.e., for 80\% of the orbit, we checked our results by calculating orbital solutions using the RVs from $\chi^2$-mapping only when the lines are unblended, i.e., near periastron, and the results of a single Gaussian fit otherwise (i.e., for $0.1<\phi<0.9$). The weights were halved when the RVs of the primary and secondary are supposed identical. These solutions are given in the last columns of Table\,\ref{SOLORB} and shown in Fig. \ref{fig:solorb} (right panel) ; they are identified by a ``prime'' sign. As could be expected, the difference between the center-of-mass velocities is reduced and the semi-amplitudes are slightly enlarged. It must however be noted that the orbital parameters are similar, within the 1-$\sigma$ error bars (except for the center-of-mass velocities, which agree within 2-$\sigma$). The largest differences are seen in the physical parameters, as the larger amplitudes naturally result in larger masses and semi-axes, hence suggesting larger inclinations ($\sim70^{\circ}$ in the case of supergiants, $\sim90^{\circ}$ for main-sequence objects).

\section{The X-ray emission from \cy}
X-ray emission from Cyg\,OB2 was discovered serendipitously when {\it EINSTEIN} was pointed at Cyg\,X-3 \citep{har79}. This was actually the first report of X-ray emission from OB stars. Since then, the region has been observed several times by X-ray observatories. The most recent exposures have been obtained thanks to a monitoring of Cyg\,OB2 \#8a with \xmm\ \citep[see preliminary results in][]{rau05}.

\subsection{The \xmm\ dataset}
In total, \xmm\ provided six pointings centered on Cyg\,OB2 \#8a (ObsId=20045, 50511, PI G. Rauw). The first four datasets, separated by 10 days each and with a duration of 20\,ks, were obtained in October - November 2004; the last two datasets, of length 30\,ks, were taken three years later, around 2007 May 1 (Table \ref{xmm}). All exposures were obtained with the same EPIC configuration (full frame, medium filter). The reduction process of the first four datasets is explained in detail in \citet{deb06}. The last two were reduced in a similar manner and we only repeat here the most important information. 

The raw data were processed with SAS version 6.0 package. Some bad time intervals affected by high background events (so-called soft-proton flares) were rejected. A few stray-light features (due to singly reflected photons) from Cyg\,X-3 are visible in the lower right corner of the images. However, they do not affect the most interesting part of the field of view. The \cy\ data were extracted within a circle (of radius 40'' for the first four datasets, of radius 30'' for MOS and 23.25'' for pn in the last two observations) centered on the source, whereas the background was extracted from a nearby 50''$\times$20'' area devoid of X-ray sources. The EPIC spectra were analysed with the {\tt xspec}v11.2 software. Note that the EPIC-pn data of \cy\ in Obs. \#4 and \#6 are not available due to the source falling partially or totally in a CCD gap.

\subsection{X-ray properties}

\begin{figure}
  \includegraphics[width=9.5cm]{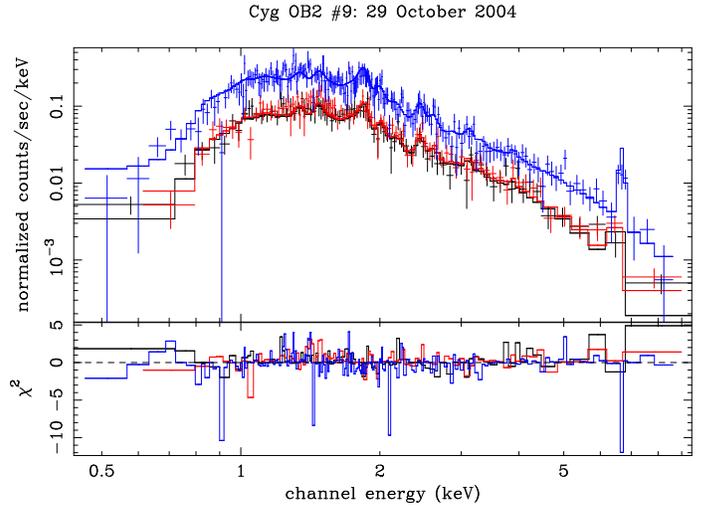}
  \includegraphics[width=8cm]{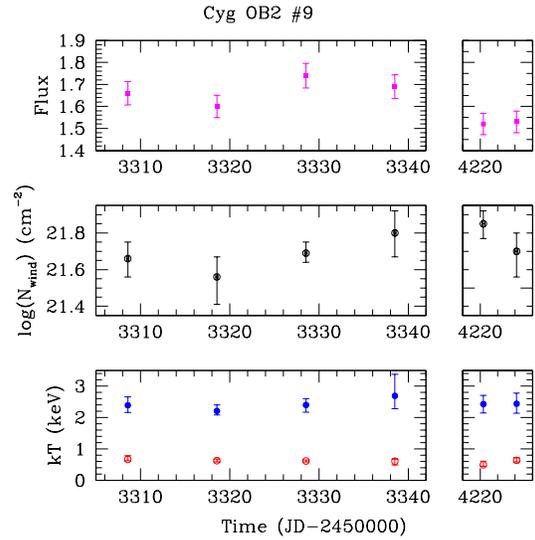}
  \hfill
  \caption{Left: EPIC spectra of \cy\ and best-fit model in the first observation (lower lines: EPIC-MOS1 in black + MOS2 in red ; upper line: pn in blue). Right: Variations of the fitted parameters through the observations.}
  \label{fig:epic}
\end{figure}

In the radio range, \cy\ displays a clear signature of non-thermal emission. Indeed, considering its quite long period, \cy\ fits in the so-called standard scheme for colliding-wind massive binaries accelerating particles, and producing synchrotron radiation in the radio domain \citep{deb07}. However, its X-ray emission, revealed by our \xmm\ observations, appears clearly thermal in nature, as several X-ray lines are detected in the spectra (e.g., the iron line at 6.7\,keV, see Fig. \ref{fig:epic}). In fact, this strong thermal X-ray emission could easily hide, in the 1--10.\,keV range, a faint putative non-thermal X-ray component due to inverse Compton scattering. To be unveiled, such an X-ray emission should be investigated using observatories with a large sensitivity above 10\,keV, where the thermal emission becomes negligible \citep{deb09}.

To obtain a good fit to the observed \xmm\ spectra, the combination of two hot, optically-thin plasma was needed, and the fitted models are thus of the type $wabs_{\rm int} \times abs_{\rm wind} \times (mekal_1+mekal_2)$. The first absorption corresponds to the interstellar one, fixed to N$_{\rm H, ISM} = 1.15 \times 10^{22}$\,cm$^{-2}$. This value was derived from the reddening of \citet{mas91} and the gas-to-dust ratio of \citet{boh78}. To allow for additional, circumstellar absorption, a second absorbing component was added and allowed to vary. As the circumstellar material actually is an ionized wind, we used the dedicated opacity tables from the wind absorption model of \citet{naz04}, as for Cygnus\,OB\,\#8a in \citet{deb06}. Table \ref{xmmfit} lists the best-fit models obtained for the six pointings. Note that f$_X^{\rm corr}$ corresponds to the dereddened flux, i.e., the flux corrected for the interstellar absorption.

The overall luminosity is rather typical of O-stars. Using the bolometric correction of \citet{mar05} for an O5I star (the combined type of \cy) as well as the V magnitude and reddening from \citet{mas91}, the $\log (L_{\rm X}/L_{\rm BOL}$) is found to be $-6.3$, close to the typical value of this ratio for O-type stars \citep[$-6.45$ with a dispersion of 0.51 dex in the 2XMM survey][]{naz09}. The average temperature ($<kT>=\sum (kT_i\times norm_i)/\sum(norm_i)$) derived from the fits is 1.2\,keV, which is slightly high compared to the average temperatures derived in the 2XMM (where 83\% of the objects have $<kT>$ below 1\,keV). In fact, when accounting for the severe ISM absorption, the spectral shape appears similar to that derived in the 2XMM for HD\,168112, another non-thermal radio emitter, at its lowest luminosity \citep{naz09}. \cy\ is thus only slightly harder and slightly more luminous than ``normal'' stars. However, it must be kept in mind that (1) with its long period, the \cy\ binary is wide, and therefore the winds are quite diluted before colliding ; (2) the observations were taken rather far away from the periastron passage, at phases $\phi$=0.14--0.22 (Fig. \ref{fig:RV}). Additional observations during such an event are needed before one can totally exclude a significant contribution from wind-wind collision to the X-ray emission of \cy.

Finally, it must be noted that \cy\ presents some variations between our observations. While the overall spectral shape changes only slightly, the flux clearly decreased in the last two observations by about 10\%, and there seems to be some shorter-term variations of lower amplitude between the first four observations. The origin of such variations is unknown. The reason for the long-term 10\%\ change cannot be constrained before securing periastron observations, but it could simply be linked to a brightening associated with strong wind-wind collisions around periastron (cf. the cases of WR140 and $\eta$\,Car).

\section{Conclusion}

\cy\ is a rare case of non-thermal (radio) emission associated with O-stars. Such emission is now thought to be associated with wind-wind collision in a binary, and we showed two years ago that \cy\ was indeed a multiple system. However, understanding non-thermal emission requires modelling the winds in detail, which in turn requests more than a ``simple'' binarity detection.

The continuous monitoring of \cy\ has led us to the derivation of a first orbital solution. The period is long, 2.4yrs, in agreement with the observed long-term radio modulation; the system eccentricity is large, 0.7--0.75, while the mass ratio approaches unity: \cy\ is thus also one of the few known long-period O+OB binaries and one of the few O+OB systems presenting a high eccentricity. It should be noted that the RV curve of \cy\ is peculiar, with only one unblended configuration (blueshifted primary - redshifted secondary) seen at periastron. Near apastron, the small RV difference between the two components is compatible with the large eccentricity and the orientation of the orbit, but the absence of RV crossing (or a limited one) requires in addition a large difference in the center-of-mass velocities for the two components. 

An additional monitoring was performed at high energies, with the hope of finding a signature of a wind-wind collision (which should be the origin of the non-thermal radio emission). In the X-ray range, however, \cy\ displays no large overluminosity nor any strong enhancement of its hard emission. There is thus, at least outside periastron, no clear, unquestionable signature of X-ray emission from the wind-wind collision. However, the flux varies, on both short and longer-timescales, with a 10\%\ brightness decrease between 2004 and 2007. The cause of these variations needs to be be investigated, notably by getting data closer to periastron.

The next periastron passage of \cy\ should occur in June-July 2011, this time without any solar conjunction problem. It is the best time for finishing the study of this object by performing a multi-wavelength campaign (simultaneous radio, X-rays, and optical monitoring), which will finally open the possibility of modelling the rare high-energy phenomena occurring in this system.

\acknowledgments
We thank E. Gosset for his advices on orbital solutions and statistics. The Li\`ege group also acknowledges financial support from the FRS/FNRS (Belgium), as well as through the Gaia-DPAC and XMM+INTEGRAL PRODEX contracts (Belspo), the ``Action de Recherche Concertée'' (CFWB-Académie Wallonie Europe), the Scientific Cooperation program 2005-2006 between Italy and the Belgian ``Communaut\'e Fran\c caise'' (project 05.02), the European Community's Seventh Framework Programme (FP7/2007-2013) under grant agreement number RG226604 (OPTICON), and the ``Cr\'edit d'impulsion ULg no. I-06/13'' (ULg). Their travels to OHP were supported by the ``Communaut\'e Fran\c caise'' (Belgium). Finally, P. Eenens thanks Conacyt for its support.

\clearpage

\clearpage

\clearpage

\clearpage

   \begin{table}
\begin{center}
      \caption{Journal of observations. Heliocentric Julian dates (mean values if N$\neq$1) are given in the format HJD--2\,450\,000, N is the number of spectra taken, $\Delta\lambda$ is the wavelength range, R is the resolving power ($\lambda/FWHM_{calib}$), S/N is the average signal-to-noise ratio of the individual exposures around 5835\AA\ (or 4555\AA\ for the blue spectrum). }
         \label{journ}
         \begin{tabular}{lccccc}
\tableline\tableline
Instrument &  Date & N & $\Delta\lambda$ (\AA)& R & S/N \\
            \tableline
Espresso  & 3288.806& 1 & 3780-6950 &18000 & 100\\
Aur\'elie & 4560.563& 3 & 5500-5900 & 8800 & 40 \\
          & 4591.557& 3 & 5500-5900 & 8800 & 80 \\
          & 4620.472& 3 & 5500-5900 & 8800 & 70 \\
          & 4652.481& 3 & 5500-5900 & 8800 & 70 \\
          & 4680.444& 4 & 5500-5900 & 8800 & 60 \\
          & 4711.439& 1 & 4450-4900 & 7000 & 30 \\
          & 4715.388& 3 & 5500-5900 & 8800 & 75 \\
WIRO      & 4672.764& 4 & 5300-6700 & 4000 & 200\\
          & 4746.743& 1 & 5300-6700 & 4000 & 150\\
          & 4818.624& 1 & 5300-6700 & 4000 & 220\\
          & 4842.070& 2 & 5300-6700 & 4000 & 150\\
Sophie    & 4906.675& 3 & 3900-6900 & 35000& 65 \\
          & 4936.968& 3 & 3900-6900 & 35000& 80 \\
          & 4953.595& 1 & 3900-6900 & 35000& 75 \\
          & 5006.497& 2 & 3900-6900 & 35000& 80 \\
          & 5022.578& 2 & 3900-6900 & 35000& 80 \\
          & 5058.433& 1 & 3900-6900 & 35000& 75 \\
          & 5079.965& 2 & 3900-6900 & 35000& 70 \\
          & 5123.370& 1 & 3900-6900 & 35000& 75 \\
          & 5142.240& 1 & 3900-6900 & 35000& 95 \\
            \tableline
         \end{tabular}
\end{center}
   \end{table}

\clearpage
   \begin{table}
\begin{center}
      \caption{RVs of the \hei\ line at 5875.62\AA, with heliocentric Julian dates given in the format HJD--2\,450\,000. The RVs have been corrected for remaining shifts using the DIB at 5780\AA\ (see text). Phases correspond to the SB2 solution (2nd column of Table \ref{SOLORB}).}
         \label{RV}
         \begin{tabular}{lcrrr}
            \tableline\tableline
Date & $\phi_{SB2}$ & $RV_1$(\kms) & $RV_2$(\kms) & $RV_{\rm Gauss}$(\kms)\\
            \tableline
3288.806 & 0.12& -60.3  & 2.5    & -41.4 \\ 	
3726.231 & 0.63& -52.6  & 15.9   & -22.6 \\ 	
3887.493 & 0.82& -47    & 34	 & -7.   \\     	
3990.455 & 0.94& -78.1  & 70.2   &       \\ 	
4022.252 & 0.98& -142.5 & 96.1   &       \\ 	
4036.237 & 1.00& -119.9 & 130.9  &       \\ 	
4051.389 & 0.02& -79.2  & 112.7  &       \\ 	
4066.298 & 0.03& -108   & 19.6   &       \\ 	
4244.476 & 0.24& -24.2  & -4.9   & -16.7 \\ 	
4303.495 & 0.31& -27.9  & 1	 & -21.3 \\     	
4324.427 & 0.34& -36    & -0.4   & -19.6 \\ 	
4348.765 & 0.37& -26.6  & -0.4   & -17.3 \\ 	
4379.063 & 0.40& -25.9  & -10.1  & -17.1 \\ 	
4463.748 & 0.50& -22    & 1.8    & -10.1 \\ 	
4472.247 & 0.51& -24.7  & -1	 & -17.8 \\     	
4560.563 & 0.61& -29.1  & 26.2   & -18.1 \\ 	
4591.557 & 0.65& -18.8  & 23.8   & 8.	 \\ 	
4620.472 & 0.68& -23.9  & 23.6   & 14.2  \\ 	
4652.481 & 0.72& -21.8  & 19.8   & -1.   \\ 	
4672.764 & 0.75& -26.3  & 11.5   & -2.7  \\ 	
4680.444 & 0.76& -22.4  & 29.8   & 13.	 \\ 	
4715.388 & 0.80& -11.4  & 23	 & 1.3	 \\      	
4746.743 & 0.83& -17.5  &  24.2  & 0.7 	 \\  	
4818.624 & 0.92& -30.8  &  52.4  &       \\  	
4842.070 & 0.95& -48.4  &  71.1  &       \\  	
4906.675 & 0.02& -97.1  &  64.5  &       \\  	
4936.968 & 0.06& -68.9  &  17.3  &       \\  	
4953.595 & 0.08& -54.7  &  7.5   &       \\  	
5006.497 & 0.14& -40.3  &  3.8   & -24.3 \\  	
5022.578 & 0.16& -36.5  & -10.9  & -25.9 \\  	
5058.433 & 0.20& -27.4  & -8.2   & -15.9 \\  	
5079.965 & 0.22& -32.7  & -6.8   & -17.2 \\  	
5123.370 & 0.28& -26.3  & -15.2  & -20.  \\  	
5142.240 & 0.30& -28.2  & -10.7  & -19.1 \\  	
            \tableline
         \end{tabular}
\end{center}
   \end{table}

\clearpage
   \begin{sidewaystable}
\begin{center}
      \caption{Preliminary orbital solution for \cy. SB1' and SB2' refer to the use of identical velocities when the lines are blended (see text).}
         \label{SOLORB}
         \begin{tabular}{lcccccc}
            \tableline\tableline
Parameter & SB2 & SB1 for Prim.& SB1 for Sec.& SB1 for ($RV_1-RV_2$)& SB1' for ($RV_1-RV_2$) & SB2'\\
            \tableline
$P$(d)                    & 852.9$\pm$4.3  & 858.6$\pm$7.0  & 848.4$\pm$3.0  & 851.4$\pm$5.5  & 851.3$\pm$5.6  & 852.8$\pm$4.4  \\ 
$T_0$                     &  4036.8$\pm$3.6&  4019.3$\pm$5.7&  4047.6$\pm$2.3&  4036.5$\pm$4.4&  4031.0$\pm$4.9&  4030.9$\pm$3.9\\  
$e$                       & 0.744$\pm$0.030& 0.752$\pm$0.033& 0.799$\pm$0.033& 0.736$\pm$0.036& 0.704$\pm$0.034& 0.708$\pm$0.027\\  
$\omega$($^{\circ}$)      &$-$164.4$\pm$4.1& 167.1$\pm$5.8  & 33.5$\pm$4.8   &$-$166.5$\pm$5.3&$-$176.4$\pm$5.9&$-$175.1$\pm$4.4\\
$M_1/M_2$                 & 1.17$\pm$0.22  &                &                & 1.27$\pm$0.35  & 1.16$\pm$0.26  & 1.10$\pm$0.17  \\ 
$\gamma_1$(\kms)          & $-$40.6$\pm$3.2& $-$38.5$\pm$2.0&                & $-$38.9$\pm$4.0& $-$24.4$\pm$5.3& $-$28.1$\pm$3.4\\ 
$\gamma_2$(\kms)          & 16.6$\pm$3.5   &                & 18.9$\pm$2.0   & 17.8$\pm$4.0   & 2.2$\pm$5.3    & 1.2$\pm$3.5    \\
$K_1$(\kms)               & 53.0$\pm$7.0   & 59.6$\pm$6.5   &                & 50.3$\pm$11.0  & 62.3$\pm$10.7  & 63.2$\pm$6.8   \\
$K_2$(\kms)               & 62.1$\pm$8.1   &                & 73.2$\pm$7.7   & 63.9$\pm$11.0  & 72.0$\pm$10.7  & 69.4$\pm$7.4   \\
$a_1\sin i$(R$_{\odot}$)  & 598.0$\pm$84.4 &                &                & 572.9$\pm$129.6& 744.3$\pm$132.7& 752.1$\pm$86.0 \\
$a_2\sin i$(R$_{\odot}$)  & 699.3$\pm$97.7 &                &                & 727.8$\pm$132.2& 860.2$\pm$134.3& 825.9$\pm$93.7 \\ 
$M_1\sin^3 i$(M$_{\odot}$)& 21.7$\pm$7.2   &                &                & 22.8$\pm$10.2  & 41.0$\pm$15.4  & 38.0$\pm$10.1  \\ 
$M_2\sin^3 i$(M$_{\odot}$)& 18.6$\pm$6.1   &                &                & 17.9$\pm$8.7   & 35.5$\pm$14.0  & 34.6$\pm$9.1   \\ 
rms(\kms)                 & 14.5           & 11.6           & 8.1            & 14.0           & 12.5           & 16.3           \\  	
            \tableline
         \end{tabular}
\end{center}
   \end{sidewaystable}

\clearpage
   \begin{table}
\begin{center}
      \caption{Journal of the \xmm\ observations, with heliocentric Julian dates given in the format HJD--2\,450\,000 and phases from the SB2 solution (2nd column of Table \ref{SOLORB}). Xspec count rates (in ct\,s$^{-1}$) within the extraction region are given for the three instruments in the 0.4--10.\,keV energy band.}
         \label{xmm}
         \begin{tabular}{llcccc}
            \tableline\tableline
Obs & Date & $\phi$ & MOS1 & MOS2 & pn\\
            \tableline
1 & 3308.583& 0.14& 0.148$\pm$0.004 & 0.162$\pm$0.004 & 0.406$\pm$0.007 \\
2 & 3318.558& 0.16& 0.138$\pm$0.004 & 0.151$\pm$0.004 & 0.397$\pm$0.008 \\
3 & 3328.543& 0.17& 0.135$\pm$0.004 & 0.141$\pm$0.004 & 0.303$\pm$0.005 \\
4 & 3338.505& 0.18& 0.131$\pm$0.004 & 0.143$\pm$0.005 &                 \\
5 & 4220.355& 0.21& 0.108$\pm$0.004 & 0.100$\pm$0.004 & 0.301$\pm$0.007 \\
6 & 4224.170& 0.22& 0.108$\pm$0.004 & 0.109$\pm$0.003 &                 \\
           \tableline
         \end{tabular}
\end{center}
   \end{table}

\begin{table}
\begin{center}
\caption{Best-fit parameters models for \cy\ and X-ray fluxes at Earth in the 0.5--10.\,keV band.}
         \label{xmmfit}
\begin{tabular}{c c c c c c c c c}
\tableline\tableline
Obs & $\log{{\rm N}_{\rm wind}}$ & kT$_1$ & norm$_1$ & kT$_2$ & norm$_2$ & $\chi^2_{\nu}$ (d.o.f.) & f$_X^{\rm obs}$ & f$_X^{\rm corr}$ \\
& cm$^{-2}$& keV & $10^{-3}$\,cm$^{-5}$& keV & $10^{-3}$\,cm$^{-5}$& & erg\,cm$^{-2}$\,s$^{-1}$ & erg\,cm$^{-2}$\,s$^{-1}$ \\
\tableline
\vspace*{-3mm}\\
1 & $21.66^{+.09}_{-.10}$ & $0.67^{+.12}_{-.04}$ & $(3.41^{+0.83}_{-0.81}) $ & $2.39^{+.27}_{-.23}$ & $(1.89^{+.28}_{-.18}) $ & 0.86 (391) & $1.66 \times 10^{-12}$ & $5.35 \times 10^{-12}$ \\
\vspace*{-3mm}\\
2 & $21.56^{+.11}_{-.15}$ & $0.63^{+.05}_{-.04}$ & $(2.58^{+0.70}_{-0.64}) $ & $2.21^{+.19}_{-.13}$ & $(2.10^{+.16}_{-.18}) $ & 0.89 (423) & $1.60 \times 10^{-12}$ & $5.24 \times 10^{-12}$ \\
\vspace*{-3mm}\\
3 & $21.69^{+.06}_{-.05}$ & $0.62^{+.03}_{-.03}$ & $(4.21^{+0.68}_{-0.87}) $ & $2.40^{+.20}_{-.23}$ & $(1.95^{+.25}_{-.17}) $ & 1.16 (414) & $1.74 \times 10^{-12}$ & $5.91 \times 10^{-12}$ \\
\vspace*{-3mm}\\
4 & $21.80^{+.12}_{-.13}$ & $0.61^{+.07}_{-.12}$ & $(4.49^{+3.07}_{-1.36}) $ & $2.69^{+.69}_{-.41}$ & $(1.80^{+.34}_{-.37}) $ & 0.95 (122) & $1.69 \times 10^{-12}$ & $5.04 \times 10^{-12}$ \\
\vspace*{-3mm}\\
5 & $21.85^{+.07}_{-.08}$ & $0.50^{+.11}_{-.05}$ & $(5.45^{+1.88}_{-2.19}) $ & $2.43^{+.28}_{-.28}$ & $(1.92^{+.18}_{-.22}) $ & 0.94 (211) & $1.52 \times 10^{-12}$ & $4.73 \times 10^{-12}$ \\
\vspace*{-3mm}\\
6 & $21.70^{+.10}_{-.14}$ & $0.64^{+.08}_{-.05}$ & $(3.10^{+1.01}_{-0.96}) $ & $2.44^{+.34}_{-.30}$ & $(1.80^{+.27}_{-.24}) $ & 0.79 (155) & $1.53 \times 10^{-12}$ & $4.67 \times 10^{-12}$ \\
\vspace*{-3mm}\\
\tableline
\end{tabular}
\end{center}
\end{table}

\end{document}